\documentclass{iopart} 

\usepackage{iopams}
\usepackage{setstack}
\usepackage{graphicx}

\newcommand{\eq}[1]{\begin{equation}#1\end{equation}}
\newcommand{\eqmulti}[1]{\begin{eqnarray}#1\end{eqnarray}}
\newcommand{\eqref}[1]{\eref{#1}}

\newcommand{\bra}[1]{{\big<{#1}\big|\,}}
\newcommand{\ket}[1]{{\,\big|{#1}\big>}}
\newcommand{\matrixe}[3]{{\big<{#1}\big|\,{#2}\,\big|{#3}\big>}}

\newcommand{\rV}{{\vec{r}}}
\newcommand{\xV}{{\vec{x}}}
\newcommand{\EC}{{\mathcal{E}}}
\newcommand{\dd}{{\mathrm{d}}}

\begin{document}


\letter{Stability of Trapped Ultracold Fermi Gases\\
  Using Effective s- and p-Wave Contact-Interactions}

\author{R. Roth and H. Feldmeier}

\address{Gesellschaft f\"ur Schwerionenforschung (GSI)\\
  Planckstr. 1, 64291 Darmstadt, Germany}

\ead{\mailto{r.roth@gsi.de}, \mailto{h.feldmeier@gsi.de}}

\begin{abstract}   
The stability of trapped dilute Fermi gases against collapse towards
large densities is studied. A hermitian effective contact-interaction
for all partial waves is derived, which is particularly suited for a
mean-field description of these systems. Including the s- and p-wave
parts explicit stability conditions and critical particle numbers are
given as function of the scattering lengths. The p-wave contribution
determines the stability of a single-component gas and can
substantially modify the stability of a two-component gas. Moreover it
may give rise to a novel p-wave stabilized high-density phase.
\end{abstract}


\maketitle
 

Since the first realization of a Bose-Einstein condensate of
${}^{87}$Rb atoms in 1995 \cite{AnEn95} the field of trapped
ultracold atomic gases experienced great experimental progress.
This raised the question if a Fermi gas can be prepared under similar
conditions and whether a transition to a superfluid state can be achieved.
An important step towards a possible super-fluid state of
trapped Fermi gases was the cooling of ${}^{40}$K atoms to a
temperature regime where degeneracy dominates \cite{DeJi99}.

A major problem for the evaporative cooling of fermions is the Pauli
exclusion principle. The relative wave-function of two
indistinguishable fermions has odd parity and hence they do not feel
the s-wave part of the interaction which dominates the force between
bosons at low kinetic energies. This limits the efficiency of
evaporative cooling. Several techniques are under discussion to
circumvent this problem, e.g., simultaneous trapping of two fermionic
species \cite{DeJi99,HoDe00}, sympathetic cooling of a fermion-boson
mixture \cite{NyMo99}, and the use of p-wave resonances to enable
efficient cooling via p-wave interactions \cite{DeBo99,MaYo98,Bohn00}.

In this letter we investigate the question how p-wave interactions
influence the properties of degenerate Fermi gases. The Thomas-Fermi
approximation together with a new hermitian effective
contact-interaction is employed to describe the effects of the
atom-atom interaction in a dilute gas including s- and p-wave
contributions.  We use this formalism to investigate the influence of
p-wave interactions on the stability of trapped one- and two-component
Fermi gases against collapse towards high densities where the atoms
escape from the trap due to three-body collisions and the formation of
bound dimers.


In the standard description of trapped atomic gases one uses point
interactions in the s-wave channel only. In the following we present a
hermitian effective contact-interaction (ECI) for all partial waves,
which is derived to be an effective mean-field interaction.

Consider two particles of mass $m$ which interact via a spherical
potential with a range much smaller than their relative wave
length. An auxiliary boundary condition at a large radius leads to a
discrete energy spectrum $\bar{E}_{nl}$, where $n$ numbers the
positive energy eigenstates and $l$ denotes the relative angular
momentum. Negative energy states, i.e., bound states, need not be
taken into account since the ECI is used for the description of
systems that are not self-bound. The energy shift $\Delta
E_{nl}=\bar{E}_{nl}-E_{nl}$ with respect to the eigenvalues
$E_{nl}=k^2_{nl}/m$ of the kinetic energy without interaction can be
expressed in terms of the phase shift $\eta_l(k)$ for
the $l$-th partial wave \cite{RoFe00}.

For mean-field type calculations we require that the expectation value
of the ECI calculated with mean-field states, i.e., free two-body states
$\ket{nlm_l}$, equals the exact energy shift 
\eq{ \label{eq:energycond}
\matrixe{nlm_l}{v^{\mathrm{eff}}_l}{nlm_l}
\overset{!}{=}
\Delta E_{nl} .
}
We choose the following hermitian ansatz for the operator of the ECI
in a particular $l$-channel
\eq{ \label{eq:veff}
v^{\mathrm{eff}}_l
= \int\!\dd^3r\;\ket{\rV}
  \frac{\overset{\leftarrowtail}{\partial}{}^l}{\partial r^l}\;\;
  g_l\;\frac{\delta(r)}{4\pi r^2}\;
  \frac{\overset{\rightarrowtail}{\partial}{}^l}{\partial r^l}
  \bra{\rV} . 
}
The arrows above the derivatives indicate to which side they act.
Using this ansatz in condition \eqref{eq:energycond} we obtain an
explicit expression for the interaction strength $g_l$ 
\eq{ \label{eq:interaction_strength}
g_l
= -\frac{4\pi}{m} \bigg[\frac{(2l+1)!!}{l!}\bigg]^{\!2}
  \frac{\eta_l(k)}{k^{2l+1}}  
\approx \frac{4\pi}{m}\frac{(2l+1)}{(l!)^2}\; a_l^{2l+1} ,
}
where $a_l$ is the scattering length of the $l$-th partial wave. It
turns out that \eqref{eq:interaction_strength} does not depend
explicitly on the auxiliary boundary condition and hence we
generalized the result to continuous momenta and
energies. Approximation of the phase shifts $\eta_l(k)$ in terms of
the scattering lengths $a_l$, so that $g_l$ is momentum independent,
leads to deviations of at most 5\% in the energy shift up to $|k
a_l|\sim3$. For the s-wave and p-wave component this reduces to $g_0 =
4\pi a_0/m$ and $g_1 = 12\pi a_1^3/m$, respectively.

The s-wave part is identical to the widely used local contact
interaction. Another way to include higher partial wave terms is the
pseudo-potential of K. Huang and C.N. Yang \cite{Huan63,HuYa57}. This
approach leads to non-hermitian interactions for $l>0$.  The two-body
energy shift induced by this pseudo-potential is by a factor
$\frac{l+1}{2l+1}$ smaller than the exact one, and is therefore not
suited for mean-field calculations beyond s-waves.


Using the s- and p-wave contributions of the ECI we calculate the
Hartree-Fock energy-density functional of interacting Fermi gases
composed of $\Xi$ distinguishable components in an external potential
$U(\xV)$. We use the Thomas-Fermi approximation, which is excellent for
these systems \cite{HoFe97}. The local energy density of the Fermi gas
as a function of the local Fermi momenta
$\kappa_{\xi}(\xV)=[6\pi^2\rho_{\xi}(\xV)]^{1/3}$ of the different
components $\xi = 1,\dots,\Xi$ reads
\eqmulti{ \label{eq:energydens}
\fl \EC[\kappa_1(\xV),\dots,\kappa_{\Xi}(\xV)] 
&= \frac{1}{6\pi^2}\sum_{\xi} U_{\xi}(\xV)\, \kappa_{\xi}^3(\xV)
  + \beta_t\sum_{\xi} \kappa_{\xi}^5(\xV) 
  + \beta_0\sum_{\xi'>\xi} 
    \kappa_{\xi}^3(\xV)\,\kappa_{\xi'}^3(\xV)  \nonumber \\ 
&+ \beta_1\sum_{\xi} \kappa_{\xi}^8(\xV) 
  + \beta_1\sum_{\xi'>\xi}\!\frac{1}{2}\big[
    \kappa_{\xi}^3(\xV)\,\kappa_{\xi'}^5(\xV) +
    \kappa_{\xi}^5(\xV)\,\kappa_{\xi'}^3(\xV) \big], 
}
with coefficients 
\eq{ \label{eq:energydens_const}
\beta_{t} = \frac{1}{20 \pi^2 m} \;,\quad
\beta_0 = \frac{a_0}{9\pi^3\,m} \;,\quad
\beta_1 = \frac{a_1^3}{30\pi^3\,m}\;.
}
The first two terms are caused by the external potential and the
kinetic energy, respectively. The remaining ones are the contributions
of the s- and p-wave parts of the ECI. As a direct consequence of the
Pauli principle, fermions of one kind interact only via the p-wave
part, while fermions belonging to different components feel s- and
p-wave interactions. The coefficient of the last term may be modified
by effective range corrections of the s-wave channel. Since in the
following applications $a_1$ is treated as a parameter this effect and
others which contribute to the same order in $\kappa_{\xi}$ are
absorbed in $a_1$.

The ground state properties of the system are determined by
minimization of the energy with respect to the local Fermi momentum
under the constraint of a fixed particle number $N_{\xi}$ for each
component. Implementing the constraints by means of chemical
potentials $\mu_{\xi}$ the variational problem reduces to an algebraic
equation for the local Fermi momentum.



First we consider a single-component Fermi gas, where the extremum
condition is given by
\eq{ \label{eq:1spec_extremalcond}
m[\mu-U(\xV)] = 
  \frac{1}{2}\,\kappa^2(\xV)  
  + \frac{8}{15\pi}\,a_1^3\;\kappa^5(\xV) .  
}
Usually the s-wave contribution dominates at low energies, but since
s-wave scattering is excluded here only p-wave scattering contributes
to the mean-field energy.


From the extremum condition \eqref{eq:1spec_extremalcond} we get the
following upper bound for the local Fermi momentum of a
single-component gas
\eq{ \label{eq:1spec_stabilitycond_kF}
-a_1\,\kappa(\xV) \le \frac{(3\pi)^{1/3}}{2},
}
beyond which no energy minimum exists anymore. Condition
\eqref{eq:1spec_stabilitycond_kF} can also be expressed as an upper
bound for the chemical potential
\eq{ \label{eq:1spec_stabilitycond_mu}
a_1^2\,m[\mu-U(\xV)] \le \frac{3(3\pi)^{3/2}}{40} .
}
This limits the particle number $N$ of a stable single-component
condensate for attractive p-wave interactions. 


The stability condition \eqref{eq:1spec_stabilitycond_kF} involves the
local Fermi momentum or density at some point in space, which is not
easy to access experimentally. Therefore, we express the stability
condition by a phenomenological parameterization as function of
particle number, scattering length and trap size. For simplicity we
restrict ourselves to an external potential $U(\xV)$ with the shape of a
deformed harmonic oscillator, where $\ell = (\ell_x\ell_y\ell_z)^{1/3}
=(m^3\omega_x\omega_y\omega_z)^{-1/6}$ is the geometric mean of the
oscillator lengths. The parameterization is fitted to the critical
particle numbers obtained from the numerical solution of the extremum
condition \eqref{eq:1spec_extremalcond} for a given scattering length
using the maximum chemical potential \eqref{eq:1spec_stabilitycond_mu}
for a stable condensate. We find that the parameterized stability
condition
\eq{ \label{eq:1spec_stabilitycond}
C\;\Big(\!\sqrt[6]{\!N}\frac{a_1}{\ell}\Big) \le 1 
\quad\mathrm{with} \quad C = -2.246,
}
for the single-component gas reproduces the numerically calculated
stability limit within errors of 1\%.

The critical particle number $N_{\mathrm{c}}=[\ell/(Ca_1)]^6$ of a
single-component gas for a typical set of experimental parameters
($a_1 = -200 a_{\mathrm{B}}$, $\ell = 1\mu\mathrm{m}$) is about
$8\times10^9$. This is much larger than the particle numbers achieved
in present experiments. Nevertheless, this stability limit could be
exceeded in near future, e.g., by increased interaction strength or by
the use of tightly confining traps \cite{ViGi00}.



In the following we consider a gas composed of two fermionic species
residing in two magnetic substates of hyperfine splitting. Although
the s-wave interaction between the species usually dominates, we show
that p-wave interactions cannot be neglected in general.

The energy minimization using the energy-density functional
\eqref{eq:energydens} for a two-component system leads to two coupled
equations for $\kappa_1(\xV)$ and $\kappa_2(\xV)$
\eqmulti{ \label{eq:2spec_extremalcond_full}
m[\mu_1-U_1(\xV)] 
&= \frac{1}{2}\,\kappa_1^2(\xV) 
   + \frac{2}{3\pi}\,a_0\;\kappa_2^3(\xV) \nonumber \\
&+ \frac{1}{30\pi}\,a_1^3 \big[ 16\,\kappa_1^5(\xV)
   + 3\,\kappa_2^5(\xV) 
   + 5\,\kappa_1^2(\xV)\kappa_2^3(\xV) \big] .
}
The second equation is generated by the exchange
$[\mu_1-U_1(\xV)]\leftrightarrow[\mu_2-U_2(\xV)]$ and
$\kappa_1(\xV)\leftrightarrow\kappa_2(\xV)$. These two equations have
a great variety of possible solutions, depending on the signs and the
relative strength of $a_0$ and $a_1$. In this letter we restrict
ourselves to equal particle numbers or $[\mu-U(\xV)] =
[\mu_1-U_1(\xV)] = [\mu_2-U_2(\xV)]$ which leads to equal local Fermi
momenta $\kappa(\xV)=\kappa_1(\xV)=\kappa_2(\xV)$ and the single
extremum condition for both components:
\eqmulti{ \label{eq:2spec_extremalcond}
m[\mu-U(\xV)]  
  = \frac{1}{2}\kappa^2(\xV)  
  + \frac{2}{3\pi}a_0\,\kappa^3(\xV) 
  + \frac{4}{5\pi}a_1^3\,\kappa^5(\xV) .
}
The generalization to different chemical potentials, different
interaction strengths or more components is straightforward.


For the two-component system with s-wave interactions only, the
stability condition was studied in \cite{HoFe97}. Using
\eqref{eq:2spec_extremalcond} which includes the p-wave interaction we
obtain the more general condition
\eq{ \label{eq:2spec_stabilitycond_kF}
-a_0\kappa(\xV) - 2\,(a_1\kappa(\xV))^3 \le \frac{\pi}{2} 
}
for each component not to collapse just because of mean-field
effects. For $a_1=0$ this reduces to the stability condition given in
\cite{HoFe97}, while for $a_1<0$ the condition limits the particle
number of a metastable two-component fermion condensate even more. It
also shows, that a repulsive s-wave interaction ($a_0>0$) does not
guarantee a stable condensate at larger densities if an attractive
p-wave component is present. In this context it is also interesting to
study the transition to spatially separated components.

\begin{figure}
\begin{center}
\includegraphics[width=0.81\textwidth]{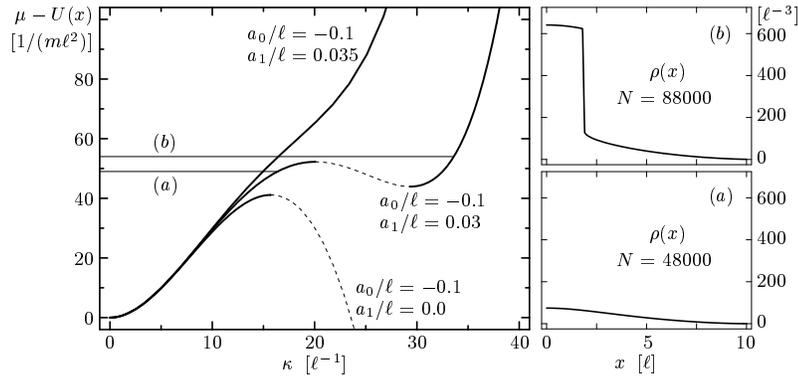}
\caption{R.h.s. of equation \eqref{eq:2spec_extremalcond} as function
  of the local Fermi momentum $\kappa$. The three curves correspond to
  different interaction parameters as labeled. Solutions of
  \eqref{eq:2spec_extremalcond} on the dotted parts of the
  curves correspond to maxima of the energy density, while the full
  branches denote (local) minima. The inserts (a) and
  (b) show the densities for two values of $\mu$ indicated by the thin
  lines for a harmonic trap with mean oscillator length $\ell$.}
\label{fig:2spec_solutionstruc}
\end{center}
\end{figure}

Several very interesting effects appear for interactions with
attractive s-wave ($a_0<0$) and repulsive p-wave components
($a_1>0$). Here $a_1$ need not be the true p-wave scattering length
but may include effective range and other effects which contribute to
the $\kappa^8$-order in the energy density. In figure
\ref{fig:2spec_solutionstruc} the r.h.s. of the extremum condition
\eqref{eq:2spec_extremalcond} is depicted as function of the local
Fermi momentum $\kappa$. For an attractive pure s-wave interaction
(lower curve) the r.h.s. exhibits an absolute maximum which leads to
an upper bound for the chemical potential as discussed for the
single-component gas. If a repulsive p-wave contribution is added,
then the r.h.s. of \eqref{eq:2spec_extremalcond} grows at large Fermi
momenta due to the leading $\kappa^5$ contribution. For values of the
s- and p-wave scattering lengths which satisfy the condition
\eq{ \label{eq:2spec_nolimitcond}
\frac{a_1}{|a_0|} \ge \frac{2}{3\pi^{2/3}} \approx 0.311 
}
the stability condition \eqref{eq:2spec_stabilitycond_kF} is fulfilled
for all values of the local Fermi momentum, i.e., the r.h.s. of the
extremum condition is a monotonic function of $\kappa$ and the maximum
does not appear anymore (upper curve). Thus the repulsive p-wave
interaction stabilizes the condensate for any particle number even
though the s-wave interaction is attractive.

If the ratio of the scattering lengths is below the limit given by
\eqref{eq:2spec_nolimitcond}, then the r.h.s. of the extremum
condition exhibits a low and a high density branch, as shown by the
middle curve of figure \ref{fig:2spec_solutionstruc}. For chemical
potentials $\mu$ below the value of the local maximum one obtains the
regular low-density solution everywhere in the potential $U(\xV)$.  An
example is shown in insert (a) of figure
\ref{fig:2spec_solutionstruc}. For larger $\mu$ novel high-density
solutions exist in areas of the trap where $m[\mu-U(\xV)]$ is larger than
the value of the local maximum. As depicted in insert (b) the density
shows a discontinuous jump from the outer low-density to the central
high-density phase. This phase is stabilized only due to the presence
of the repulsive p-wave interaction and occurs when the condition
\eqref{eq:2spec_stabilitycond_kF} is violated. For values of $\mu$ in
between the maximum and the minimum a Maxwell construction with equal
pressure in the high- and low-density regime may be used to identify
the equilibrium density.

However, when $a_1/|a_0|$ drops below $2/(3\pi^{3/2})$ the density in
the high-density regime grows rapidly so that the three-body
recombination rate and thus the trap loss is not small anymore. If
$a_1/|a_0|<\sqrt[3]{160/(729 \pi^2)}\approx 0.281$ the local minimum
of the r.h.s. of \eqref{eq:2spec_extremalcond} occurs at negative
values, so that even a self-bound solution exists. In this case the
product $|a_0|\,\kappa(\xV=0) > \frac{9}{8}\pi$, which means that in
the self-bound area in the center of the trap the mean distance
$\rho^{-1/3}$ between the atoms gets close to the scattering length
$|a_0|$ and the underlying approximations break down. Nevertheless we
expect that for interactions with $a_1/|a_0|\approx0.3$ a metastable
high-density phase occurs, where the central densities are by one
order of magnitude higher than in the outer low-density region.


Like for the single-component gas we express the stability condition
in terms of the particle number $N$ of each component, the mean
oscillator length $\ell$ of a harmonic trap, and the scattering
lengths $a_0$ and $a_1$. We use the following slightly more
complicated parameterization for the stability condition
\eq{ \label{eq:2spec_stabilitycond}
C_0 \Big(\!\sqrt[6]{\!N}\frac{a_0}{\ell}\Big)
+ C_1^3 \Big(\!\sqrt[6]{\!N}\frac{a_1}{\ell}\Big)^{\!3} 
+ C_{01}^{n+1} \Big(\!\sqrt[6]{\!N}\frac{a_0}{\ell}\Big) 
  \Big(\!\sqrt[6]{\!N}\frac{a_1}{\ell}\Big)^{\!n} \le 1 .
}
The fit has to be done separately for each type of attractive
interaction. The resulting optimal coefficients are summarized in table
\ref{tab:2spec_coefficients} and reproduce the numerical stability
limit again with deviations of less than 1\%.
\begin{table}[b]
\caption{Parameters of the fitted stability condition 
  \eqref{eq:2spec_stabilitycond} for the two
  component condensate for different interaction types.}
\label{tab:2spec_coefficients}
\begin{indented}
\item[]
\begin{tabular}{c c c c c}
\br
interaction type & $C_0$ & $C_1$ & $C_{01}$ & $n$ \\
\mr
$a_0\le0,\,a_1\le0$ & -1.835 & -2.570 & 0.656  & 1  \\
$a_0\ge0,\,a_1<0$   & -1.378 & -2.570 & 1.360  & 1  \\
$a_0<0,\,a_1\ge0$   & -1.835 & -1.940 & 2.246  & 3  \\
\br
\end{tabular}
\end{indented}
\end{table}

\begin{figure}
\begin{center}
\includegraphics[width=0.6\textwidth]{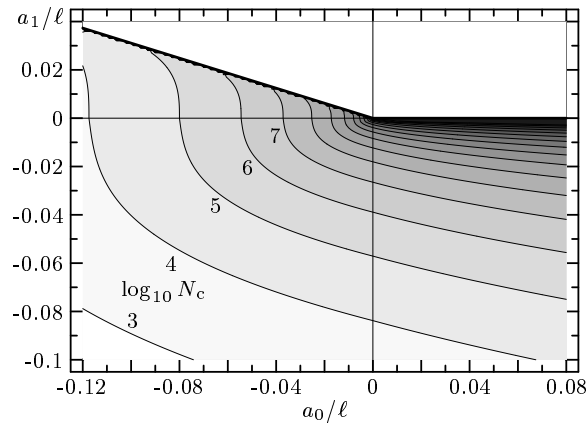}
\caption{Logarithmic contour plot of the critical particle
  number $N_{\mathrm{c}}$ as function of the s- and p-wave scattering 
  length for the two-component fermion condensate. Neighboring contours are
  separated by a factor 10 in particle number, selected ones are
  labeled with $\log_{10} N_{\mathrm{c}}$. The white area indicates the
  parameter region where the particle number is not limited.}
\label{fig:2spec_stabmap}
\end{center}
\end{figure}

The critical particle numbers $N_{\mathrm{c}}$ resulting from the
stability condition \eqref{eq:2spec_stabilitycond} are shown in figure
\ref{fig:2spec_stabmap} as function of $a_0$ and $a_1$. For negative
p-wave scattering length $N_{\mathrm{c}}$ describes the maximum particle
number of a metastable condensate. The white area in the figure marks
the region where from the mean-field point of view a metastable
condensate with smooth density exists for all particle numbers. 
In principle a positive $a_1$ always leads to a stable mean-field
solution. But for $a_0<0$ and values of $a_1/|a_0|\lesssim0.28$ this
solution is at a density which is too large for the metastable
state. Therefore, the limit for $N_{\mathrm{c}}$ resulting from
\eqref{eq:2spec_stabilitycond} is plotted. The bending over of the
contour lines close to $a_1/|a_0|\approx0.3$ indicates the onset of
the novel high-density phase.

As an example for estimating from figure \ref{fig:2spec_stabmap} the
critical particle number we take a two-component ${}^6$Li condensate
in a trap with $\ell=1\mu m$. The extraordinary large attractive
s-wave scattering length of $a_0 = -2160 a_{\mathrm{B}}$ \cite{AbMc97},
i.e. $a_0/\ell\approx-0.1$, leads to the rather small critical
particle number $N_{\mathrm{c}} = 26000$ for each of the two
components. As also seen from figure \ref{fig:2spec_stabmap} the p-wave
interaction will change $N_{\mathrm{c}}$ significantly whenever
$|a_1|/|a_0|\gtrsim0.2$.

We conclude that for Fermi gases p-wave scattering should not be
neglected from the outset. If attractive it constrains the total
particle number in the metastable condensate, if repulsive it helps to
stabilize multi-component systems with attractive s-wave
interactions. In special cases even a novel high-density phase may
occur in the center of the trap which should not be confused with Bose
condensation of Cooper pairs.




\section*{References}

\begin{thebibliography}{10}

\bibitem{AnEn95}
M. Anderson {\it et~al.}, Science {\bf 269},  198  (1995).

\bibitem{DeJi99}
B. DeMarco and D.~S. Jin, Science {\bf 285},  1703  (1999).

\bibitem{HoDe00}
M.~J. Holland, B. DeMarco, and D.~S. Jin, Phys. Rev. A {\bf 61}, 053610 (2000).

\bibitem{NyMo99}
N. Nygaard and K. M{\o}lmer, Phys. Rev. A {\bf 59},  2974  (1999).

\bibitem{DeBo99}
B. DeMarco {\it et~al.}, Phys. Rev. Lett. {\bf 82},  4208  (1999).

\bibitem{MaYo98}
M. Marinescu and L.You, Phys. Rev. Lett. {\bf 81},  4596  (1998).

\bibitem{Bohn00}
J.~L. Bohn, Phys. Rev. A {\bf 61}, 053409 (2000).

\bibitem{RoFe00}
R. Roth and H. Feldmeier, GSI Annual Report 1999 {\bf GSI 2000-01}, (2000), a
  detailed publication on the effective contact interaction is in preparation.

\bibitem{Huan63}
K. Huang, {\em Statistical Mechanics} (John Wiley \& Sons, New York, 1963),
  Chap.~13, p.\ 274.

\bibitem{HuYa57}
K. Huang and C. Yang, Phys. Rev. {\bf 105},  767  (1957).

\bibitem{HoFe97}
M. Houbiers {\it et~al.}, Phys. Rev. A {\bf 56},  4864  (1997), and references
  therein.

\bibitem{ViGi00}
L. Viverit {\it et~al.}, e-print cond-mat/0005517.

\bibitem{AbMc97}
E.~R.~I. Abraham {\it et~al.}, Phys. Rev. A {\bf 55},  R3299  (1997).

\end{thebibliography}
\end{document}